\documentclass[11pt,fleqn,twoside]{article}
\usepackage{amsfonts,amssymb,latexsym}
\makeatletter
\newcommand{\prava}{\footnotesize\it
\begin{flushright}
\begin{minipage}{18cm}
Copyright \copyright 1998 by Q.J.A. Khan, B.S. Bhatt and R.P. Jaju
\end{minipage}
\end{flushright}}

\newcommand{\name}[1]{\begin{flushleft}
                       \LARGE \bf #1
                       \end{flushleft}\vspace{-3mm}}

\newcommand{\Author}[1]{\begin{flushleft}
                       \it #1 \end{flushleft}}

\newcommand{\Adress}[1]{\begin{flushleft}
                       \it #1 \end{flushleft}}

\newcommand{\Date}[1]{\begin{flushleft}
                      \small  \it #1 \end{flushleft}}

\newcommand{\ehkol}{Author \ name}
\newcommand{\ohkol}{Article \ name}
\renewcommand{\@evenhead}{
\hspace*{-3pt}\raisebox{-15pt}[\headheight][0pt]{\vbox{\hbox to \textwidth
{\thepage \hfil \ehkol}\vskip4pt \hrule}}}
\renewcommand{\@oddhead}{
\hspace*{-3pt}\raisebox{-15pt}[\headheight][0pt]{\vbox{\hbox to \textwidth
{\ohkol \hfil \thepage}\vskip4pt\hrule}}}
\renewcommand{\@evenfoot}{}
\renewcommand{\@oddfoot}{}

\newcommand{\be}{\begin{equation}}
\newcommand{\ee}{\end{equation}}
\newcommand{\ba}{\hspace*{-5pt}\begin{array}}
\newcommand{\ea}{\end{array}}

\newcommand{\ds}{\displaystyle}
\makeatother

\begin{document}

\thispagestyle{empty}

\renewcommand{\ehkol}{Q.J.A. Khan,  B.S. Bhatt and R.P. Jaju}
\renewcommand{\ohkol}{Switching Model with Two Habitats and Group Defence}

\begin{flushleft}
\footnotesize \sf
Journal of Nonlinear Mathematical Physics \qquad 1998, V.5, N~2,
\pageref{khan-fp}--\pageref{khan-lp}.
\hfill {\sc Article}
\end{flushleft}

\vspace{-5mm}

\renewcommand{\footnoterule}{}
{\renewcommand{\thefootnote}{}
 \footnote{\prava}}

\name{Switching Model with Two Habitats and \\
a Predator Involving Group Defence}\label{khan-fp}

\Author{Q.J.A. KHAN~$^\dag$,  B.S. BHATT~$^\ddag$ and
R.P. JAJU~$^\star$}

\Adress{$^\dag$~Department of Mathematics {\rm \&} Statistics,
Box 36, Science College\\
~~Sultan Qaboos University, PC-123, Oman\\[1mm]
$^\ddag$~Department of Mathematics {\rm \&} Computer Science\\
~~University of West Indies, Trinidad, West Indies\\[1mm]
$^\star$~Department of Computer Science, Box 36, Science College\\
~~Sultan Qaboos University, PC-123, Oman}

\Date{Received February 17, 1998; Accepted  March 13, 1998}

\begin{abstract}
\noindent
 Switching model with one predator and two prey species is considered.
The prey species have the ability of group defence. Therefore, the predator
will be attracted towards that habitat where prey are less in number. The
stability analysis is carried out for two equilibrium values. The
theoretical results are compared with the numerical results
for a set of values. The Hopf bifuracation analysis is done to support
the stability results.
\end{abstract}

\section*{1. Introduction}
Among related herbivore species, individuals of smaller species like Dik-dik
will be vulnerable to a greater range of predator species and are less
likely than larger to be able to defend themselves against, or to out
run, predators.  All small species avoid being detected by predators.
Smaller species are likely to have to seek carefully their scare, scattered
 food item of high quality and form less cohesive and coordinated feeding
 groups.  They live singly or in pairs and f\/ind their resources within a
defended territory.  Because they are small and vulnerable, they move and
feed cautiously and slowly and never move far from cover.  The size of
the territory is presumably determined by the area that a pair can defend
and by the availability of suitable food at the season of greatest scarcity.
These species characteristically remain in one vegetation type in all
seasons.

Larger species individuals feeding upon abundant, evenly dispersed, easily
found items, are likely to be tolerating low quality food.  They form
enormous, rather formless, feeding aggregations of many thousands of
animals.  Major predators of zebra, buf\/falo, kongoni, toki and Thomson's
gazelle are hyena, wild dog, lion, leopard and cheetas.  They form groups
for defence against predators and more likely depend upon self-defence,
group defence, group alertness within a group and speed, to avoid being
killed by a predator.  Dense vegetation and broken terrain disrupt visual
communication, and f\/lat open country favour it.  So such groups are more
likely to be found where visual communication is favoured and where
individuals can conform to the group, speed, and direction of movement.
Unless the group remains cohesive and coordinated, the individual risks
becoming an outstanding target.  At all times individuals in groups must
remain in communication and their speeds and directions when moving must
vary little between individuals.  Group defence is a term used to describe a
phenomenon whereby predation is decreased or even prevented altogether by
the ability of the prey population to better defend or disguise themselves
when their number is large.  Aggregation tends to reduce the risk of
predation through simple dilution.  Hence, doubling the local density of
herbivores while predator density remains unchanged could lead to a halving
of the mortality risk Hamilton [1], Bertram [2].  Of course, this assumes
that predators do not seek out areas with very large prey density Schaller
[3].  Pairs of musk-oxen can be successfully attacked by wolves but groups
are rarely attacked Tener [4].  There are many examples of group defence --
Yang and Humphrey [5], May and Robinson [6], Holmes and Bethel [7].  Herds
remain well coordinated even under attack and individuals may benef\/it from
the alertness and communication within the herd.  Individuals tend to
conform with their neighbour activities, and many hundreds, even thousands
of wildebeest can coordinate rapidly in response to an alarm.
 Large groups also benef\/it from increased probability of detection of
predators.  The hunting success of lions decline if the group size of prey
is large Van Orsdol [8].  Cheetah prefer to hunt single animals.  Coursing
predators select less healthy, old, sick and young prey and those who have
lost their herds during migrations due to various reasons.  Animals in poor
condition and without group may reduce vigilance rates.

Each year, some one million wildebeest migrate across the Serengeti mara
ecosystem D.~Kreulen [9].  The crude cost of this movement, relative to
neighbouring resident populations of wildebeest, is a 3\% increment in
mortality per year Sinclair [10].  The overall migratory pattern is thought
to be related to food supply, which is itself dependent on an uneven
distribution of rainfall Grzimek and Grzimek [11], Talbot and Talbot [12],
Anderson and Talbot [13], Pennywick [14], Maddock [15].  In the dry
season, they need to drink Jarman [16] and most home ranges include some
riverbank habitat, which is preferentially used during that season.  At the
start of wet season, they move away from the rivers to occupy the woodlands
where they feed predominantly on new green grass leaves, with some herbs,
many of these plants may be annuals.  As grasses mature and rains cease,
their diet starts to include progressively more browse and they move
into plant communities where browse is more.  Movements could also be
inf\/luenced by change in requirements for specif\/ic nutrients.  Kreulen [17]
noted that Serengeti wildebeest on their wet season range preferred a
short-grass over a long grass habitat, and that calcium concentration were
higher on the short grassland.  Migration has also been attributed to the
wildebeest's dislike of wet and sticky ground Talbot and Talbot [12],
Anderson and Talbot [13].  In this way herbivores maximize the growth
potential of the vegetation through rotational grazing where the two
concentration areas are suf\/f\/iciently far apart, the movements are called
seasonal migrations.

In predator-prey environment, the predator prefers to feed itself in a
habitat for some duration and then changes its preference to another
habitat.  This preferential phenomenon of change of habitat by the predator
is called switching.  There may be several reasons of switching of
predators.  For example -- predator prefers to catch prey species in a
habitat where they are in abundance but after some duration of heavy
predation, when the prey species population starts declining, the predator
changes its preference to another habitat.  In this habitat prey species are
more in number due to less predation i.e. the predator feeds preferentially
on the most numerous prey species habitat.  This is found to be the case
when prey species is relatively smaller in size with little or insignif\/icant
defence capability with respect to predator, like small antelope and
cruising predators to locate prey.  Many examples may be cited where a
predator prefers to prey species that is most abundant at any time, see
Fisher-Piette [18], Lawton et al. [19] and Murdoch [20].  The mathematical
models which have been generally proposed with such type of predator
switching are those involving one predator with two prey species, e.g.,
Holling [21], Takashi~[22], May [23], Murdoch and Oaten [24], and
Raughgarden and Feldman [25], Tanskey [26, 27], Prajneshu and Holgate [28],
Teramoto et al [29].

We are motivated by Serengeti ecosystem which can be divided into two
habitats for wild life -- (a) open southern grasslands with low rainfall that
support a relatively low biomass of short-growing grasses and (b) wooded
northern grasslands with higher rainfall that support tall, highly lignif\/ied
grasses Braun [30], McNaughton [31, 32].  Rainfall is a key factor
inf\/luencing primary productivity in both grassland types Braun [30] Sinclair
[33]. McNaughton [31, 32] . All of the migratory species (wildebeest, zebra,
and Thomson's gazelle) show similar seasonal shifts in habitats, using short
grassland in the south during wet season and tall grasslands in the north
during dry season Pennywick [14] and Maddock [15].  Their long range
movements roughly correspond to seasonal transitions.  We have considered a
system having a predator species interacting with same prey species in two
habitats.  Prey is able to migrate among two dif\/ferent habitats at some cost
to the population in the sense that the probability of survival during a
change of habitat may be less than one. The predator can feed on either
habitats.  The prey species in both habitats have
the ability of group defence but it will be ef\/fective in the habitat where
the population of prey is large.  The predator will be attracted toward
s that habitat where prey  are less in number.

Freedman and Walkowicz [34] considered a predator-prey system in which the
prey population exhibits group defence.  They described that if the carrying
capacity of the prey population is suf\/f\/iciently large and there is no mutual
interference among predators then the predator population always goes to
extinction.  Freedman and Quan [35] studied predator-prey system with third
population extinction.  Shigui Ruan and Freedman [36] analyzed group defence
in Gause-type models for three species.  They gave criteria for persistence
when there is no mutual interference and when there is mutual interference
of predators.  Freedman and Shigui Ruan [37] have discussed a three species
food chain model with group defence.  They have shown that model undergoes a
sequence of Hopf bifurcations, using the carrying capacity of the
environment as a bifurcation parameter.

This paper is organized as follows -- Model formulation is in Section 2
and equilibrium and stability is discussed in Section 3. Section 4 includes
Hopf bifurcation analysis with respect to $\delta$ (conversion rates of prey
to predator) and $\nu$ (death rate of the predator).  Bifurcation points and
ef\/fect on stability for selected numerical data values are in Section~5.
Final discussion and results are summarized in Section 6.

\setcounter{section}{2}
\setcounter{equation}{0}
\renewcommand{\theequation}{\arabic{section}.\arabic{equation}}

\section*{2. Prey-Predator model with group defence}

We consider a class of Voltera predator prey model with group defence
exhibition by the the prey of the form
\be
\ba{ll}
\ds \dot X_1 = (\alpha_1  - \epsilon_1){X_1} + \epsilon_2
p_{21} X_2 -\frac{\beta_1 X_2^2 Y}{(X_1 + X_2)}, & X_1(0) \ge 0 ,\\[4mm]
\ds \dot X_2 = (\alpha_2  - \epsilon_2) {X_2} + \epsilon_1
p_{12} X_1 - \frac{\beta_2 X_1^2Y}{(X_1 + X_2)} , & X_2 (0) \ge 0,
\\[4mm]
\ds \dot Y = \left(-\nu + \frac{\delta_1 \beta_1 X_2^2}{(X_1+
X_2)} + \frac{\delta_2 \beta_2 X_1^2}{(X_1 + X_2)}\right) Y, &
 Y(0) \ge 0,
\ea
\ee
where $\ds \left(\cdot = \frac{d}{dt}\right),$ and $X_1(t), X_2(t)$
are prey species in the f\/irst and second habitats respectively and
$Y(t)$ denote predator species which feed upon
$X_1$ and $X_2$. Prey species are able to disperse among 2 dif\/ferent
habitats at some cost to the population. Prey have the ability of
group defence, so predator
will go towards the habitats where prey population is less numerically.
We consider that
in the beginning, prey species are less in habitat $1$ and so predator
will be attracted towards them. For giving protection to their mates
the prey species of the habitat
$2$ will rush towards habitat $1$. In this way their strength in the
habitat $2$ will fall
short and the predators will be attracted towards them. So to protect
them prey start coming from
hibitats $1$ to $2$. This implies a kind of switching from one habitat
source of food
to another habitat as the prey fall short alternately. Due
to seasonal migration of prey species, none of the habitat population
will be extinct. This situation is described by the above model where
\begin{enumerate}
\item[$\beta_i:$]  Predator response rates towards the two prey $X_1$
and $X_2$ respectively.
\item[$\delta_i:$] The rate of conversion of prey to predator.
\item[$\epsilon_i:$] Inverse barrier strength in going out of the f\/irst
habitat and the second habitat.
\item[$p_{ij}:$]  The probability of successful transition from
$i^{\mbox{\scriptsize th}}$ habitat to $j^{\mbox{\scriptsize th}}$ habitat.
\item[$\alpha_i:$] Specif\/ic growth rate of the prey in the absence of
predation.
\item[$\nu:$] Per capita death rate of predator.
\end{enumerate}

We assume that $\beta_i, \delta_i,\, \epsilon_i, p_{ij},\, \alpha_{i}$,
and $\nu$ are positive constants and $\epsilon_1 p_{12}>
\beta_2Y(0),$
$\epsilon_2p_{21}>\beta_1 Y(0),$ so that the $ X_1 $ and $ X_2 $ are not
negative.

\setcounter{section}{3}
\setcounter{equation}{0}

\section*{3. Analytical Solution}
In this section, we proceed to analyze the system (2.1).
We examine the equilibrium of this system. We obtain equilibrium solutions
by setting time derivatives to zero. If in equations (2.1)
$\bar X_1,  \bar X_2, \bar Y $ are the
equilibrium values of $X_1, X_2, Y$ respectively, there are two possible
equilibria, namely,
\[
\mbox{(i)~} \quad \bar X_1 = \bar X_2 =
\bar Y = 0, \quad \mbox{i.e., population
is extinct and this always exists.}
\]
\[
\mbox{(ii)} \quad  \bar X_1 = \frac{\nu(\bar X +1)
\bar X}{(\delta_1 \beta_1 + \delta_2 \beta_2 \bar X^2)},
\qquad \bar X_2 = \frac{\nu(1 + \bar X)}
{(\delta_1 \beta_1 + \delta_2 \beta_2 \bar X^2)},
\]
\[
\phantom{\mbox{(ii)} \quad }
\bar Y = \frac{(1+\bar X)\left\{(\alpha_2-\epsilon_2)
+\epsilon_1 p_{12}\bar X\right\}}{\beta_2\bar X^2}
\qquad \mbox{or equivalently}
\]
\[
\phantom{\mbox{(ii)} \quad }
\bar Y =\frac{(1+\bar X)\left\{(\alpha_1 -\epsilon_1)
\bar X +\epsilon_2 p_{21}\right\}}{\beta_1}.
\]
Here, $\bar X = \left(\bar X_1/\bar X_2\right)$
is a real positive root of the cubic equation,
\be
\beta_2(\alpha_1 -\epsilon_1)\bar X^3 +\beta_2\epsilon_2
p_{21}\bar X^2  -\beta_1\epsilon_1 p_{12}\bar X -\beta_1
(\alpha_2 - \epsilon_2)=0.
\ee
For equilibrium values $\left(\bar X_1, \bar X_2,
\bar Y\right)$ to be positive, a positive real root of (3.1) must be
bounded as
\be
\frac{\epsilon_2 -\alpha_2}{\epsilon_1 p_{12}} < \bar
 X < \frac{\epsilon_2 p_{21}}{\epsilon_1 -\alpha_1},
 \ee
Since in the absence of the predator $\ds \frac{dX_1}{dt} >0$ and
$\ds \frac{dX_2}{dt}>0.$

From Descarte's sign rule the equation (3.1) has unique positive
root and hence the unique positive equilibrium solution of model (2.1).

\medskip

\noindent
{\bf (a) Stability Analysis of Equilibirum (i).} Consider a small
perturbation about the equilibrium level
$X_1 = \bar X_1 +u$, $X_2 = \bar X_2 +v$, $Y =
\bar Y+w.$ Substituting these into the
dif\/ferentiale equation (2.1) and neglecting products of small quantities,
 we obtain stability matrix
\be
\left(
\ba{ccc}
\alpha_1-\epsilon_1 & \epsilon_2 p_{21} & 0\\[1mm]
\epsilon_1 p_{12} & \alpha_2-\epsilon_2 & 0\\[1mm]
0 & 0 & -\nu
\ea \right).
\ee
The characteristic equation of this matrix is
\be
(\nu+\alpha) \left[\lambda^2 -\lambda \{\alpha_1-\epsilon_1)
+(\alpha_2-\epsilon_2)\}
+(\alpha_1-\epsilon_1)(\alpha_2-\epsilon_2) -\epsilon_1\epsilon_2
p_{12}p_{21}\right]=0.
\ee
If $\alpha_1 < \epsilon_1$ and $\alpha_2 <\epsilon_2$ and
$(\alpha_1 -\epsilon_1) (\alpha_2 -\epsilon_2) > \epsilon_1\epsilon_2
p_{12}\ p_{21}$, the equilibrium
$\bar X_1 = \bar X_2 = \bar Y = 0$ is locally
stable otherwise unstable.

\medskip

\noindent
{\bf (b) Stability Analysis of Equilibrium (ii).} Following the
above procedure as in (a) the stability matrix becomes:
\be
\left(
\ba{ccc}
A & \ds -\frac{\bar X_1}{\bar X_2} A & \ds -\frac{\beta_1\bar X_2^2}
{(\bar X_1 +\bar X_2)} \\[4mm]
\ds -\frac{\bar X_2}{\bar X_1} B & B & \ds -\frac{\beta_2\bar X_1^2}
{(\bar X_1+\bar X_2)} \\[4mm]
C\bar Y & D\bar Y & 0
\ea \right).
\ee
The characteristic equation associated with the positive equilibrium
$(\bar X_1, \bar X_2, \bar Y)$ of this model is
\be
\ba{l}
\ds \lambda^3 -\lambda^2 (A+B) +\lambda \left(\frac{C\beta_1\bar Y
\bar X_2^2}{(\bar X_1 +\bar X_2)}
+\frac{D\bar Y\beta_2\bar X_1^2}{(\bar X_1 +\bar X_2)}\right)\\[4mm]
\ds \qquad  -\left(\frac{AC\beta_2\bar Y\bar X_1^3}
{\bar X_2(\bar X_1 +\bar X_2)}+
\frac{BC\beta_1\bar Y\bar X_2^2}{(\bar X_1 +\bar X_2)} +\frac{AD
\bar Y \beta_2\bar X_1^2}{(\bar X_1 +\bar X_2)} +
\frac{BD\bar Y \beta_1\bar X_2^3}{\bar X_1(\bar X_1+\bar X_2)}\right)
= 0,
\ea
\ee
where
\[
\ba{l}
\ds A =(\alpha_1-\epsilon_1)+\frac{\beta_1\bar X_2^2 \bar Y}
{(\bar X_1 +\bar X_2)^2}, \qquad
B= (\alpha_2 -\epsilon_2) +\frac{\beta_2\bar X_1^2\bar Y}
{(\bar X_1+\bar X_2)^2},\\[4mm]
\ds C = \frac{(-\nu+2\bar X_1 \delta_2\beta_2)\bar Y}{(\bar X_1+\bar X_2)},
\qquad
D = \frac{(-\nu+2\bar X_2\delta_1\beta_1)\bar Y}{\bar X_1 +\bar X_2)}.
\ea
\]
Equation (3.6) can be written in the form
\be
\lambda^3+ a_1\lambda^2+a_2\lambda +a_3 = 0,
\ee
where
\be
\ba{l}
\ds a_1 = -(A+B),\qquad
 a_2 = \frac{C\beta_1\bar Y \bar X_2^2}{(\bar X_1+\bar X_2)} +
 \frac{D\bar Y\beta_2\bar X_1^2}{(\bar X_1+\bar X_2)},\\[4mm]
\ds a_3 = - \left(\frac{AC \beta_2\bar Y\bar X_1^3}{\bar X_2
(\bar X_1 +\bar X_2)} +
\frac{BC\beta_1\bar Y\bar X_2^2}{(\bar X_1 +\bar X_2)} +\frac{AD
\bar Y\beta_2\bar X_1^2}{(\bar X_1+\bar X_2)}
+\frac{BD\bar Y\beta_1 \bar X_2^3}{\bar X_1(\bar X_1+\bar X_2)}\right).
\ea
\ee
The Routh-Hurwitz stability criteria for the third order system is
\[
\ba{l}
\mbox{(a)} \quad a_1>0, \ a_3>0,\\[1mm]
\mbox{(b)} \quad a_1a_2> a_3.
\ea
\]
Hence, the equilibrium (ii) will be locally stable to small perturbations
if it satisf\/ies the following conditions
\be
\ba{l}
 A + B<0,\\[1mm]
\ds  \left(C\bar X_1 +D\bar X_2\right)
\left(A\beta_2 \bar X_1^3 +B\beta_1\bar X_2^3\right) < 0,\\[2mm]
\ds  \left(\beta_1\bar X_2^3 -\beta_2\bar X_1^3\right)
\left(BD\bar X_2 - AC\bar X_1\right) > 0.
\ea
\ee
Here we observe that the stability of the model depends upon the
conditions (3.2) and (3.9) together with various parameters.

\setcounter{section}{4}
\setcounter{equation}{0}

\section*{4. Hopf Bifurcation Analysis}

We investigate the Hopf bifurcation for the system (2.1) taking $\delta_1,
\delta_2 $ and $\nu$ as the bifurcation parameters.

First, we determine the criteria for Hopf bifurcation using $\delta_1$
(rate of conversion of prey in the habitat $1$ to predator) as the
bifurcation parameter.
For non zero equilibrium we look at the characteristic equation (3.7).

\[
\mbox{For}\quad A<0, B<0, C>0, \quad \mbox{and}\quad  D>0;
\]
$a_1, a_2$ and $a_3$ are positive in (3.8), clearly (3.7) has two purely
imaginary roots if\/f $a_1 a_2 = a_3 $ for some value of $\delta_1$
(say $\delta_1 =\bar\delta_1).\ $ There exists a unique $\bar\delta_1$
such that   $a_1a_2 = a_3. $ Therefore there is only one value of
$\delta_1$ at which we have a Hopf bifurcation. Thus in the neighbourhood
of $\bar\delta_1$ the characteristic
equation (3.7) cannot have real positive roots.
For $\delta_1 =\bar\delta_1,$ we have
\be
\left(\lambda^2+a_2\right)(\lambda +a_1)=0,
\ee
which has three roots
\[
\lambda_1 = i\sqrt{a_2}, \qquad \lambda_2 = -i\sqrt{a_2},
\qquad \lambda_3 = -a_1.
\]
The roots are in general of the form
\be
\lambda_1(\delta_1) = p(\delta_1) + iq(\delta_1),
\qquad \lambda_2(\delta_1) = p(\delta_1) -iq(\delta_1),
\qquad \lambda_3(\delta_1) = -a_1(\delta_1).
\ee
To apply Hopf's bifurcation Theorem [38] to (2.1), we have to
verify the transverality condition
\be
Re\left(\frac{d\lambda_k}{d\delta_1}\right)_{\delta_1 =\bar\delta_1}
 \ne 0,\qquad k=1, 2.
 \ee
Substituting $\lambda_k(\delta_1) = p(\delta_1) +iq(\delta_1)$ into
(3.7) and calculating the derivative, we get
\be
\ba{l}
R(\delta_1)p'(\delta_1)-S(\delta_1)q'(\delta_1)+T(\delta_1) = 0,\\[1mm]
S(\delta_1)p'(\delta_1) +R(\delta_1)q'(\delta_1) + U(\delta_1) = 0,
\ea
\ee
where
\be
\ba{l}
R(\delta_1) =3p^2(\delta_1) +2a_1(\delta_1) p(\delta_1) +
a_2(\delta_1) - 3q^2(\delta_1),\\[1mm]
 S(\delta_1) = 6p(\delta_1) q(\delta_1) +2a_1(\delta_1)
 q(\delta_1),\\[1mm]
T(\delta_1) = p^2(\delta_1) a'_1(\delta_1) + a'_2(\delta_1)p(\delta_1)
+ a'_3(\delta_1) - a'_1(\delta_1) q^2(\delta_1),\\[1mm]
 U(\delta_1) = 2p(\delta_1) q(\delta_1) a'_1(\delta_1) +a'_2(\delta_1)
 q(\delta_1).
 \ea
 \ee
If $ SU + RT\ne 0$ at $\delta_1=\bar\delta_1,$ then
\be
Re\left(\frac{d\lambda_k}{d\delta_1}\right)_{\delta_1=\bar\delta_1}
= -\frac{(SU + RT)}{2(R^2+S^2)}\bigg\vert_{\delta_1
=\bar\delta_1}\ne 0.
\ee
Now from equation (4.5)
\be
SU + RT = a_1 a'_2 - a'_3 \quad \mbox{at}\quad \delta_1 = \bar\delta_1,
\ee
where $\ds  a'_2 = \frac{da_2}{d\delta_1}$ and $\ds a'_3 =
\frac{da_3}{d\delta_1}.$
$\bar X$ is a real positive root of equation (3.1) and independent of
$\ds \delta_1\ \Rightarrow\ \frac{d\bar X}{d\delta_1} = 0.$

Substituting the values of $a_1, a'_2$ and $a'_3$ in equation (4.7),
and using
$a_1 a_2=a_3$ at $\delta_1=\bar\delta_1$ which gives
\[
\left(\beta_2\bar X^2 -\frac{\beta_1}{\bar X}\right)
\left(\frac{BD}{\bar X} - AC\right)  = 0\quad \mbox{at}\quad
\delta_1 =\bar\delta_1,
\]
i.e., $\ds \frac{BD}{AC} =\bar X$, at $\delta_1 =\bar\delta_1$
and value of $\ds \frac{d\bar X_1}{d\delta_1},$ we obtain,
\be
\ba{l}
\ds  (SU+RT)= \frac{2\bar Y^2}{(1+\bar X)^2}
 \left(\frac{\beta_1}{\bar X} -\beta_2\bar X^2\right)
B\beta_1\bar X_2\\[4mm]
\ds \qquad \times\left[\frac{\bar X_2\delta_1\beta_1-\nu+
\bar X_1\delta_2\beta_2
+\bar X(-\nu+2\bar X_1\delta_2\beta_2}{-\nu+2\bar X_1
\delta_2\beta_2}\right].
\ea
\ee
Since $C$ and $D$ are positive, the terms in square bracket are positive.
 Hence
\[
Re\left(\frac{d\lambda_k}{d\delta_1}\right)_{\delta_1=\bar\delta_1}\ne 0
\quad \mbox{if}\quad \frac{\beta_1}{\beta_2}\ne
\bar X^3,\quad k = 1, 2.
\]
and $\lambda_3 (\delta_1) = -a_1(\delta_1) \ne 0.$

We summarize the above results in the following:

\medskip

\noindent
{\bf Theorem.} {\it Suppose $\bar E(\bar X_1, \bar X_2, \bar Y)$
exists and $A<0, B<0, C>0$ and $D>0.$
The system (2.1) exhibits a Hopf's bifurcation in the first octant
leading to a family of
periodic solutions that bifurcates $\bar E$ for a suitable value of
$\delta_1$ in a neighbourhood of $\bar\delta_1$ if $\ds
\frac{\beta_1}{\beta_2} \ne\bar X^3.$}

\medskip

We can get a similar result when $\delta_2$ is taken as a bifurcation
parameter. Therefore the bifurcation points that we
obtain in Table I of section 5 are the Hopf's bifurcation points.

Now we analyze the dynamics of (2.1) with respect to $\nu$ (per
capita death rate of predator).

Similar to (4.3) we need to verify
\be
Re\left(\frac{d\lambda_k}{d\nu}\right)_{\nu =\bar\nu} = -\frac{SU +RT}
{2(R^2+S^2)} \ne 0,\qquad k=1, 2,
\ee
where $S, U, R, T$ have the similar expression as given in (4.4) but
now these are functions of $\nu$ (instead of $\delta_1).$
\[
SU + RT = a_1 a'_2 -a'_3,\quad \mbox{where}\quad a'_2=\frac{da_2}{d\nu}
\quad \mbox{and}\quad a'_3 =\frac{da_3}{d\nu}.
\]
Using the relation
\be
\frac{d\bar X_1}{d\nu} =\frac{(\bar X+1)\bar X}{(\delta_1\beta_1+
\delta_2\beta_2\bar X^2)},
\ee
we can show that
\be
\frac{dC}{d\nu} = \frac{dD}{d\nu} = 0.
\ee
For purely imaginary roots of (3.7), $a_1 a_2 = a_3\ $ at $\nu = \bar \nu,$
we get
\be
(A+B) \left(\frac{C\beta_1}{\bar X} +D\beta_2\bar X\right) =
AC\beta_2\bar X^2 +\frac{BC\beta_1}{\bar X} + AD\bar X.
\ee
Now using (4.10)--(4.12) we obtain the value of
\be
a_1 a'_2 - a'_3 = 0.
\ee
Therefore, from (4.9), we get
\[
 Re\left(\frac{d\lambda_k}{d\nu}\right)_{\nu =\bar\nu}=0,\qquad
 k = 1, 2,
 \]
i.e., if $\nu$ is a bifurcation parameter, there is no Hopf's bifurcation.
Hence with respect $\nu$ the system (2.1) is either stable or unstable,
the result which we obtain in Section 2.

\setcounter{section}{5}
\setcounter{equation}{0}

\section*{5. Numerical Solutions}

 For illustration we have seen the ef\/fects of $\beta$'s, $\delta$'s.
The behaviour of stability with respect to $\beta$'s and $\delta$'s
is given in Table I.

\medskip

\centerline{{\bf Table I}}
\centerline{$\nu = 0.01$, $\alpha_1 = 0.05$, $\alpha_2 = 0.25$,
$\epsilon_1 = 0.1$, $\epsilon_2 = 0.3$, $p_{12} = 0.5$, $p_{21} = 0.2$}

\vspace{-3mm}

$$
\ba{ccccc}
\beta_1 &\beta_2 &\delta_1/\delta_2 &\mbox{Bifurcation point}
&\mbox{stable}\\[2mm]
0.01 & 0.02 & \delta_2 = 0.3 &\delta_1 = 0.75123 & 0.75124 \le
\delta_1\le 1 \\[1mm]
0.01 & 0.02 & \delta_1 = 0.5 &\delta_2 = 0.19968 & 0\le
\delta_2\le 0.19967 \\[1mm]
0.01 & 0.03 & \delta_1 = 0.5 &\delta_2 = 0.12949 & 0\le\delta_2
\le 0.12948 \\[1mm]
0.02 & 0.01 &\delta_2 = 0.3 &\delta_1 = 0.16448 & 0\le\delta_1
\le 0.16447 \\[1mm]
0.01 & 0.01 &\delta_1 = 0.5 &\delta_2 = 0.42507 &
0\le\delta_2 < 0.42506 \\[1mm]
0.015 & 0.01 &\delta_1 = 0.5 &\delta_2 = 0.66456 & 0.66457\le\delta_2
\le 1 \\[1mm]
0.02 & 0.01 &\delta_1 = 0.5 &\delta_2 = 0.91196 & 0.91197
\le\delta_2\le 1
\ea
$$

\medskip

The last 2 columns give the values of $\delta_1/\delta_2$ at which the
model is stable or unstable. In fact fourth column gives the bifurcation
point (where the model is stable, below/above that value the model is
unstable/stable).  In the appendix we show that these bifurcation points
are in fact Hopf bifurcation points.  We have also done computations to
see the ef\/fect of $\nu$ on the stability
when $\beta_1 > \beta_2.$ It is interesting to f\/ind that the model is
either stable or unstable with
respect to $\delta 's.$ and Hopf bifurcation does not exist with respect
to~$\nu$.

One can try similar analysis with respect to other parameters also.

The set of equations given in (2.1) have
been numerically integrated for four cases given in Table II with other
parameters as given in Table I.

\medskip

\centerline{\bf Table II}

\vspace{-3mm}

$$
\ba{cccccc}
\mbox{Case} & \beta_1 &\beta_2 &\delta_1 &\delta_2
&\mbox{Stable/unstable}\\[2mm]
\mbox{(i)} & 0.01 & 0.02 & 0.95 & 0.3 &\mbox{stable} \\[1mm]
\mbox{(ii)} & 0.02 & 0.01 & 0.5 & 0.3 &\mbox{stable} \\[1mm]
\mbox{(iii)} & 0.01 & 0.02 & 0.15 & 0.3 &\mbox{unstable} \\[1mm]
\mbox{(iv)} & 0.02 & 0.01 & 0.6 & 0.3 &\mbox{unstable}
\ea
$$

\medskip

These sets were picked up while doing the computations of analytical
results in previous section where the behaviour of the model is shown in the
last column. The initial conditions used are the
corresponding equilibrium values in each case with slight perturbations.
Figures 1 to 4 give the behaviour of $X_1, X_2$ and $Y$ with respect to
$t$ in above four cases and as expected we get
stable behaviour in the f\/igures 1 and 2 and unstable behaviour in the
f\/igures 3 and 4. Figures 5 to 8 give the prey-predator dynamics
when the model is stable whereas f\/igures 9 to 12 represent the
prey-predator dynamics when the model is unstable. These
contours also support the predictions of Table II.

\section*{6.  Summary and Discussion}

  We have considered a system having  a predator species interacting with
prey species in two
habitats.  Prey is of large size and migrate between two dif\/ferent
habitats at some cost to its population in the sense that the probability
of survival during a change of habitat is less than one.  The predator
can feed on either habitats.  The prey species in both habitats have
the ability of group defence but it will be ef\/fective in the habitat
where the population of prey is large.  Due to group defence ability
of the prey, predator will select the habitat where prey might have
insuf\/f\/icient defending capability (i.e. numerically less, old, sick,
some youngs and those who might have lost their group during migration
due to various reasons).  The
stability analysis has been carried out for both zero and nonzero
equilibrium values.   Nonzero equilibrium  $\bar{X_1}$  and
$\bar{X_2}$  for prey in f\/irst and second habitats depends on the
death rate of  predators i.e. if the death rate of the predator
is high then values of  $\bar{X_1}$  and  $\bar{X_2}$  will increase
or vice versa. $\bar{X_1}$  and  $\bar{X_2}$  values will decrease
if the predator response towards both habitats increases respectively.
Nonzero equilibrium will be stable if it satisf\/ies all three conditions
of equation ${(3.9)}$.  Consider f\/irst the limit where
$\epsilon_1 \rightarrow  0$  and  $\epsilon_2 \rightarrow  0$,
there is no movement of prey in or out of both habitat and so values of
$A$  and  $B$  in ${(3.9)}_1$, will be positive and nonzero equilibrium
will become unstable.  Hence, increasing the values of  $\epsilon_1$
and $\epsilon_2$  always stabilizes.  We can conclude from ${(3.9)}_2$
that stability increases for increasing values of conversion rate of
prey species by predator.  From ${(3.9)}_3$ we see that if
$\bar{X_2}  >  \bar{X_1}$  then for stability it is more likely
$\beta_1  > \beta_2$ provided $\ds \frac {BD}{AC} > \bar{X}$ or
vice versa i.e.  if the equilibrium value of prey species in second
habitat is more than f\/irst habitat then predator will
attract towards f\/irst habitat because prey exhibits group defence
and our model predicts this behaviour.

        Hopf bifurcation analysis has been carried out for both
models with respect to $\lambda$ (conversion rate of  prey to predator)
as a parameter.  In the sense of ecology, Hopf bifurcation has helped
us in f\/inding the existence of a region of instability in the
neighbourhood of nonzero equilibrium, where prey species in both the
habitats and predator  will survive undergoing regular f\/luctuations.
However, the conditions of Hopf  bifurcation might not be satisf\/ied
due to changes in other parameters and change the steady state
or otherwise.

\label{khan-lp}

\enddocument
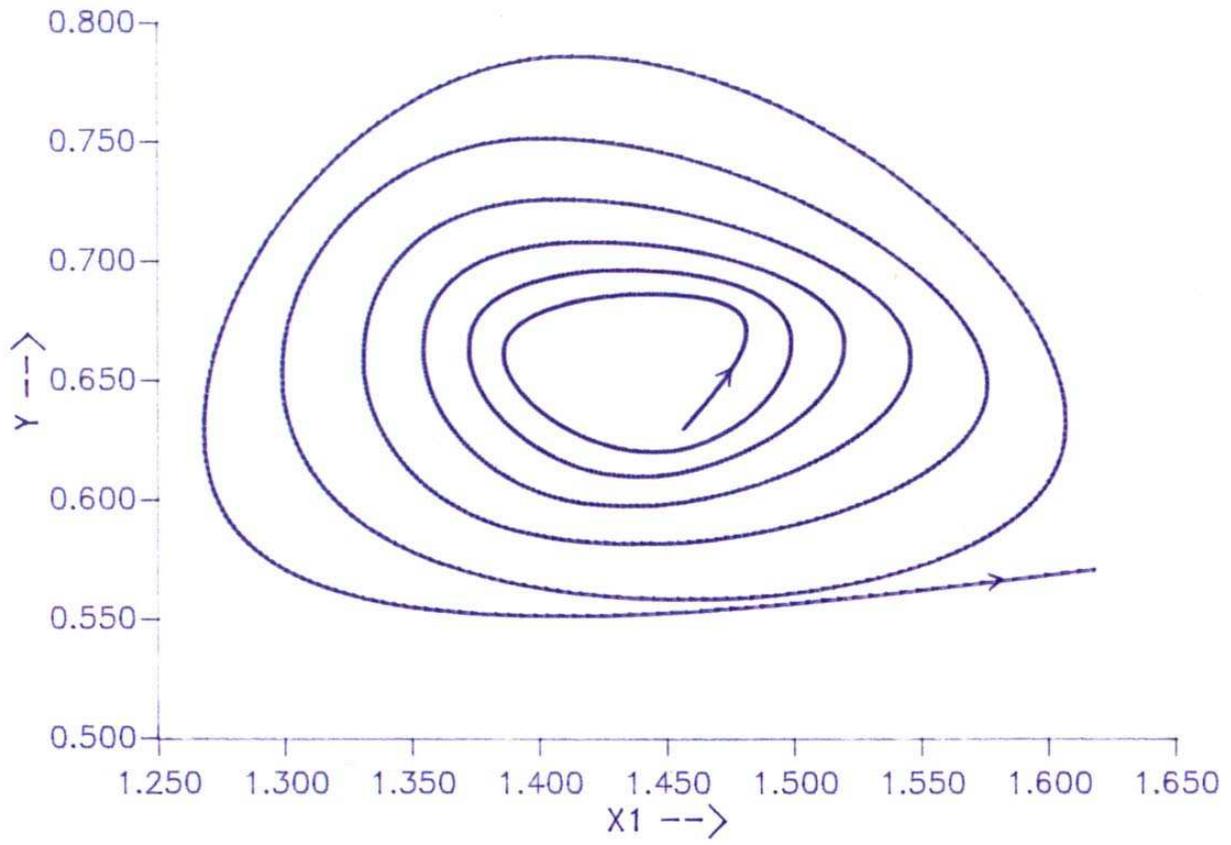
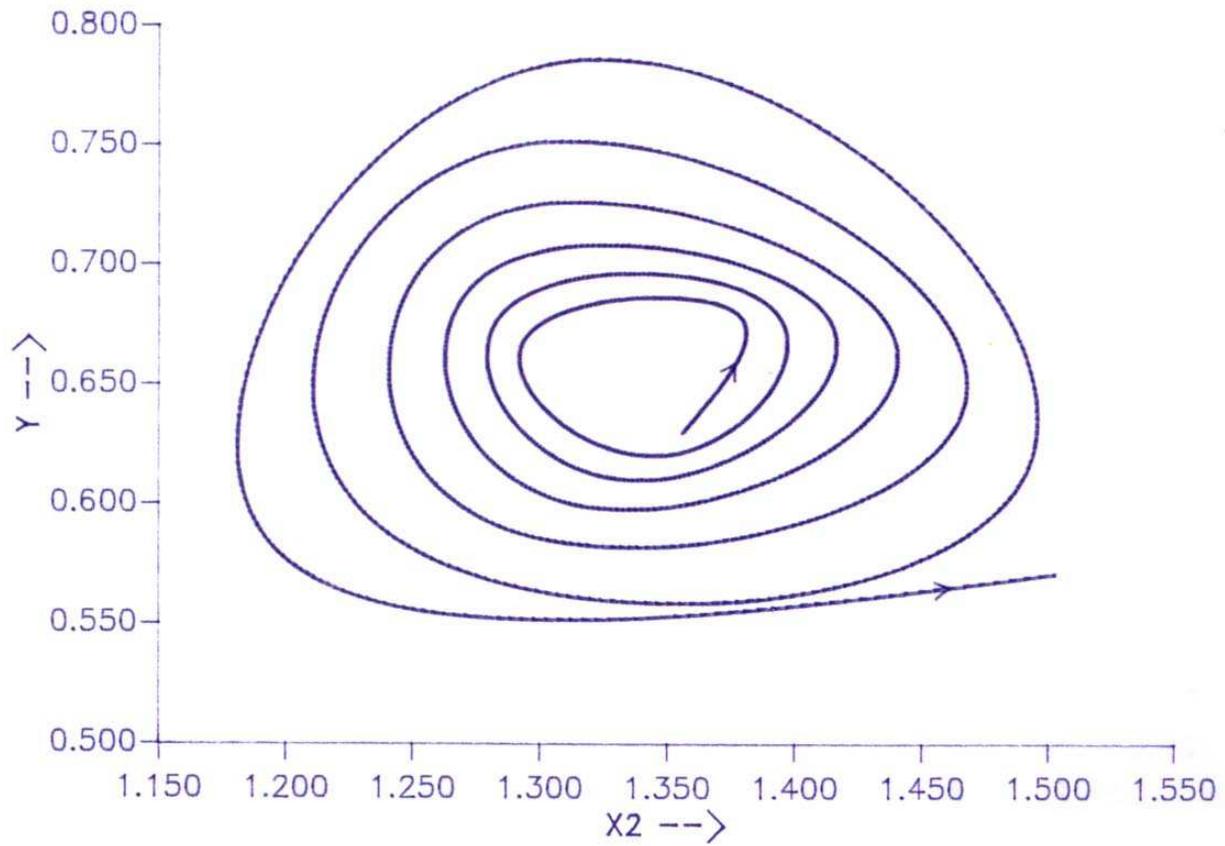
